\documentclass{pasj01}

\def\commenta{$^*$}
\def\commentb{$^\dagger$}
\def\commentc{$^\ddagger$}
\def\commentd{$^\S$}
\def\commente{$^\|$}
\def\commentf{$^\#$}

\newcounter{author}
\def\authorcount#1#2{\refstepcounter{author}\label{#1}
                     \altaffiltext{\ref{#1}}{#2}}

\begin{document}
\SetRunningHead{T. Kato et al.}{V1006 Cygni}

\Received{201X/XX/XX}
\Accepted{201X/XX/XX}

\title{V1006 Cygni: Dwarf Nova Showing Three Types
       of Outbursts and Simulating Some Features of
       the WZ~Sge-Type Behavior}

\author{Taichi~\textsc{Kato},\altaffilmark{\ref{affil:Kyoto}*}
        Elena~P.~\textsc{Pavlenko},\altaffilmark{\ref{affil:CrAO}}
        Alisa~V.~\textsc{Shchurova},\altaffilmark{\ref{affil:Kyiv}}
        Aleksei~A.~\textsc{Sosnovskij},\altaffilmark{\ref{affil:CrAO}}
        Julia~V.~\textsc{Babina},\altaffilmark{\ref{affil:CrAO}}
        Aleksei~V.~\textsc{Baklanov},\altaffilmark{\ref{affil:CrAO}}
        Sergey~Yu.~\textsc{Shugarov},\altaffilmark{\ref{affil:Sternberg}}$^,$\altaffilmark{\ref{affil:Slovak}}
        Colin~\textsc{Littlefield},\altaffilmark{\ref{affil:LCO}}
        Pavol~A.~\textsc{Dubovsky},\altaffilmark{\ref{affil:Dubovsky}}
        Igor~\textsc{Kudzej},\altaffilmark{\ref{affil:Dubovsky}}
        Roger~D.~\textsc{Pickard},\altaffilmark{\ref{affil:BAAVSS}}$^,$\altaffilmark{\ref{affil:Pickard}}
        Keisuke~\textsc{Isogai},\altaffilmark{\ref{affil:Kyoto}}
        Mariko~\textsc{Kimura},\altaffilmark{\ref{affil:Kyoto}}
        Enrique~de~\textsc{Miguel},\altaffilmark{\ref{affil:Miguel}}$^,$\altaffilmark{\ref{affil:Miguel2}}
        Tam\'as~\textsc{Tordai},\altaffilmark{\ref{affil:Polaris}}
        Drahomir~\textsc{Chochol},\altaffilmark{\ref{affil:Slovak}}
        Yutaka~\textsc{Maeda},\altaffilmark{\ref{affil:Mdy}}
        Lewis~M.~\textsc{Cook},\altaffilmark{\ref{affil:LewCook}}
        Ian~\textsc{Miller},\altaffilmark{\ref{affil:Miller}}
        Hiroshi~\textsc{Itoh},\altaffilmark{\ref{affil:Ioh}}
}

\authorcount{affil:Kyoto}{
     Department of Astronomy, Kyoto University, Kyoto 606-8502, Japan}
\email{$^*$tkato@kusastro.kyoto-u.ac.jp}

\authorcount{affil:CrAO}{
     Crimean Astrophysical Observatory, p/o Nauchny, 298409,
     Republic of Crimea}

\authorcount{affil:Kyiv}{
     Taras Shevchenko National University of Kyiv, Glushkova ave., 4, 03127,
     Kyiv, Ukraine}

\authorcount{affil:Sternberg}{
     Sternberg Astronomical Institute, Lomonosov Moscow State University, 
     Universitetsky Ave., 13, Moscow 119992, Russia}

\authorcount{affil:Slovak}{
     Astronomical Institute of the Slovak Academy of Sciences, 05960,
     Tatranska Lomnica, the Slovak Republic}

\authorcount{affil:LCO}{
     Department of Physics, University of Notre Dame, 
     225 Nieuwland Science Hall, Notre Dame, Indiana 46556, USA}

\authorcount{affil:Dubovsky}{
     Vihorlat Observatory, Mierova 4, 06601 Humenne, Slovakia}

\authorcount{affil:BAAVSS}{
     The British Astronomical Association, Variable Star Section (BAA VSS),
     Burlington House, Piccadilly, London, W1J 0DU, UK}

\authorcount{affil:Pickard}{
     3 The Birches, Shobdon, Leominster, Herefordshire, HR6 9NG, UK}

\authorcount{affil:Miguel}{
     Departamento de F\'isica Aplicada, Facultad de Ciencias
     Experimentales, Universidad de Huelva,
     21071 Huelva, Spain}

\authorcount{affil:Miguel2}{
     Center for Backyard Astrophysics, Observatorio del CIECEM,
     Parque Dunar, Matalasca\~nas, 21760 Almonte, Huelva, Spain}

\authorcount{affil:Polaris}{
     Polaris Observatory, Hungarian Astronomical Association,
     Laborc utca 2/c, 1037 Budapest, Hungary}

\authorcount{affil:Mdy}{
     Kaminishiyamamachi 12-14, Nagasaki, Nagasaki 850-0006, Japan}

\authorcount{affil:LewCook}{
     Center for Backyard Astrophysics Concord, 1730 Helix Ct. Concord,
     California 94518, USA}

\authorcount{affil:Miller}{
     Furzehill House, Ilston, Swansea, SA2 7LE, UK}

\authorcount{affil:Ioh}{
     Variable Star Observers League in Japan (VSOLJ),
     1001-105 Nishiterakata, Hachioji, Tokyo 192-0153, Japan}



\maketitle

\begin{abstract}
   We observed the 2015 July--August long outburst of V1006 Cyg
and established this object to be an SU UMa-type dwarf nova
in the period gap.  Our observations have confirmed that
V1006 Cyg is the second established object showing three
types of outbursts (normal, long normal and superoutbursts)
after TU Men.
We have succeeded in recording the growing stage of superhumps
(stage A superhumps) and obtained a mass ratio of 0.26--0.33,
which is close to the stability limit of tidal instability.
This identification of stage A superhumps demonstrated
that superhumps indeed slowly grow in systems near
the stability limit, the idea first introduced by \citet{Pdot6}.
The superoutburst showed a temporary dip followed by
a rebrightening.  The moment of the dip coincided with
the stage transition of superhumps, and we suggest that
stage C superhumps is related to the start of
the cooling wave in the accretion disk.
We interpret that the tidal instability was not strong enough to
maintain the disk in the hot state when the cooling wave
started.  We propose that the properties commonly seen
in the extreme ends of mass ratios (WZ Sge-type objects
and long-period systems) can be understood
as a result of weak tidal effect.
\end{abstract}

\section{Introduction}

   Cataclysmic variables (CVs) are composed of a white dwarf and a red
(or brown) dwarf supplying matter to the white dwarf, forming
an accretion disk.  Dwarf novae are a class of CVs characterized
by outbursts.  SU UMa-type dwarf novae are a subclass of
dwarf novae which show superoutbursts in addition to normal
outbursts.  During superoutbursts, superhumps having periods
a few percent longer than the orbital periods ($P_{\rm orb}$)
are observed and are considered to be the defining characteristics of
SU UMa-type dwarf novae [For general information of CVs,
SU UMa-type dwarf novae and superhumps, see e.g. \citet{war95book}].
The origin of superhumps and superoutbursts is currently
understood as the consequence of the 3:1 resonance in
the accretion disk resulting tidal instability combined
with thermal instability (thermal-tidal instability model;
\cite{osa89suuma}; \cite{osa13v1504cygKepler}).
Only systems having mass ratios ($q=M_2/M_1$) smaller
than $\sim$0.3 can hold the radius of the 3:1
resonance inside the tidal truncation radius (\cite{whi88tidal};
\cite{smi07SH}) and the appearance of superhumps in these
systems gave a support to the tidal instability model
for superhumps.

   In recent years, it has been established that superhump
periods during superoutbursts show systematic variations.
\citet{Pdot} showed that the evolution of superhumps
has three stages: stage A (long, constant superhump period),
stage B (short superhump period with systematic period
variations) and stage C (constant period shorter than
those of stage B superhump typically by 0.5\%; seen in the
late phase of the superoutburst to the post-superoutburst
phase).  Stage A superhumps are now considered to be
superhumps during which the 3:1 resonance is growing
and transition to stage B is considered to be caused by
the pressure effect which produces a retrograde precession
(\cite{osa13v344lyrv1504cyg}; \cite{kat13qfromstageA}).
The origin of stage C superhumps is still unknown.
It has been well established that period variations during
stage B are a good function of $P_{\rm orb}$
for systems with short $P_{\rm orb}$
(cf. \cite{Pdot}; \cite{Pdot7}).
In long-$P_{\rm orb}$ systems, however, there have been a number
of objects showing a strong decrease of the superhump
periods [the best-known examples are MN Dra and UV Gem,
see subsection 4.10 in \citet{Pdot}].  The origin of
strongly negative period derivatives had remained
a mystery.  \citet{Pdot6} proposed a working hypothesis
that what looked like strongly negative period derivatives
for stage B superhumps in such systems are actually
caused by stage A-B transition based on the photometrically
detected $P_{\rm orb}$ in MN Dra.  \citet{Pdot6} suggested
that the 3:1 resonance grows slower in systems having
large $q$ close to the stability border of the resonance
and that this is observed as long-lasting stage A.
This interpretation violated the received wisdom that
long-lasting stage A reflects the small $q$,
as is typically seen in WZ Sge-type dwarf novae
\citep{kat15wzsge}, which have completely opposite properties
to long-orbital systems.  Since the discussion in \citet{Pdot6}
was based on the yet unconfirmed $P_{\rm orb}$ of MN Dra,
further confirmation is clearly needed.  We present
the detection of long-lasting growing superhumps in
a long-$P_{\rm orb}$ system V1006 Cyg having
a spectroscopically established $P_{\rm orb}$.

   V1006 Cyg was discovered as a dwarf nova (S 7844)
(\cite{hof63v1006cyg}; \cite{hof63an287169})
with a photographic range of 16--18 mag.
\citet{ges66VS8} derived an outburst cycle length of
13.5~d.  \citet{BruchCVatlas} recorded another outburst.
\citet{bru92CVspec2} obtained a spectrum and established
this object to be a dwarf nova.  Due to the initially
reported faintness, this object had not been well
studied since then.
Since 2005, AAVSO observers started regular monitoring
using CCDs and recorded an outburst reaching $V$=13.6
on 2006 June 24.  This outburst lasted at least for 4~d.
\citet{she07CVspec} performed a radial-velocity study
and obtained $P_{\rm orb}$ of 0.09903(9)~d.
Since this period places the object in the period gap,
the slowly fading 2006 outburst was suspected to be
a superoutburst.  Upon a bright (unfiltered CCD magnitude 13.6)
outburst on 2007 August 14, a search for superhumps
was conducted (vsnet-alert 9471).  Although short-term
modulations were detected, these variations were later
found to be orbital variations (vsnet-alert 9489).
This observation and the observation of a similar bright
outburst in 2009 September (vsnet-alert 11490, 11508)
were examined in detail by \citet{pav14nyser} and
the absence of superhumps was confirmed.

\begin{figure}
  \begin{center}
    \FigureFile(70mm,88mm){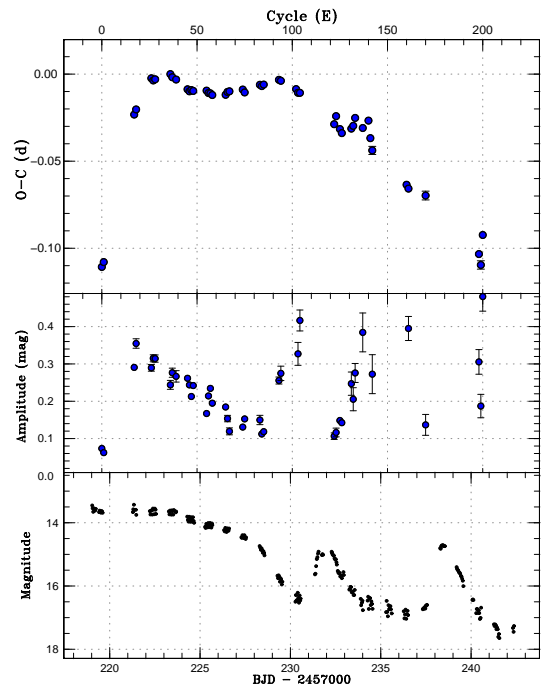}
  \end{center}
  \caption{$O-C$ diagram of superhumps in V1006 Cyg (2015).
     (Upper:) $O-C$ diagram.
     We used a period of 0.10541~d for calculating the $O-C$ residuals.
     The superhump maxima up to $E=28$ are stage A superhumps, maxima
     between $E$=36 and $E=94$ have a positive period derivative
     and are identified as stage B superhumps.  After this,
     the period decreased to a constant one (stage C superhumps).
     (Middle:) Amplitudes of superhumps.  The amplitudes were
     small around $E=0$.  The $O-C$ diagram suggests that stage A-B
     transition occurred somewhere
     between $E=28$ and $E=36$.  The superhump amplitudes monotonously
     decreased during the superoutburst.  After $E=100$,
     the amplitudes became large (0.3--0.4 mag)
     when the object faded.
     (Lower:) Light curve.  The data were binned to 0.035~d.
     The initial outburst detection was on BJD 2457215.9,
     3~d before the start of our observation.  It took 6~d
     for this object to fully develop stage B superhumps.
     The maximum on BJD 2457232 is a rebrightening.  The maximum
     on BJD 2457238 is the first normal outburst of regular
     series of outbursts following a superoutburst.
  }
  \label{fig:v1006cyghumpall}
\end{figure}

   On 2015 July 12, ASAS-SN (\cite{ASASSN}; Danilet et al.
in preparation) detected an outburst at $V$=14.1
(vsnet-alert 18846).  This outburst was detected sufficiently
early and the initial phase of the outburst was observed.
The object reached $V$=13.6 and low-amplitude superhumps
were detected on July 15--16 (vsnet-alert 18851).
The superhumps were observed to glow until July 18
(BJD 2457222) (see figure \ref{fig:v1006cyghumpall}).

\section{Observation and Analysis}\label{sec:obs}

   The data were obtained under campaigns led by 
the VSNET Collaboration \citep{VSNET}.
We also used the public data from 
the AAVSO International Database\footnote{
   $<$http://www.aavso.org/data-download$>$.
}.
Time-resolved observations were performed in 13 different
locations by using 30cm-class telescopes (table 2).
We deal with observations until August 7.
The data analysis was performed just in the same way described
in \citet{Pdot} and \citet{Pdot6} and we mainly used
R software\footnote{
   The R Foundation for Statistical Computing:\\
   $<$http://cran.r-project.org/$>$.
} for data analysis.
In de-trending the data, we divided the data into
four segments in relation to the outburst phase and
used locally-weighted polynomial regression 
(LOWESS: \cite{LOWESS}) except the rising segment
of a rebrightening.  During the rising phase of
the rebrightening a third order polynomial fitting was
instead used.
The times of superhumps maxima were determined by
the template fitting method as described in \citet{Pdot}.
The times of all observations are expressed in 
barycentric Julian Days (BJD).

\section{Discussion}\label{sec:discussion}

\subsection{Identification of superhump stages}\label{sec:superhump}

   The amplitudes of superhumps before BJD 2457220 were
small, indicating that we recorded the growing stage
(stage A) of superhumps.  Although the observations
before BJD 2457220 were short and the initial night
suffered from poor conditions, observations
between BJD 2457221 and 2457223 were of sufficient quality
to determine the period in the early phase (table 3).
The $O-C$ analysis (17$\le E \le$28) and PDM analysis
yielded periods of 0.1073(2)~d and 0.1076(1)~d,
respectively.  The period of 0.1075~d (average of the two
methods) is 8.5\% longer than $P_{\rm orb}$,
giving an exceptionally large fractional superhump excess.
Since the amplitudes of superhumps on BJD 2457221 already
approached the maximum, these periods are likely shorter than
the true period of stage A superhumps, because
the pressure effect starts to dominate when superhumps
fully grow and reduces
the precession rate \citep{kat13qfromstageA}.
By using this period as an approximate period of
stage A superhumps and with the strong expectation that
the true period of stage A superhumps is longer
than this period, we have been able to resolve
the ambiguity in the cycle counts between
BJD 2457220 and 2457221.  The cycle counts in table E2
are based on this identification.
The resultant mean period of
stage A superhumps between BJD 2457219 and 2457223
by the $O-C$ analysis is 0.1093(3)~d, which we consider
the best period from the present observations.
This period gives $\epsilon^*$ of 0.094(3), which
correspond to $q$=0.34(2).

   The duration of stage A was at least 32 cycles.
Although the true duration of the growing phase of 
superhumps is unknown
in this object due to the observational gap, the close
similarity of the $O-C$ diagram and variation of
superhumps amplitudes between V1006 Cyg and MN Dra
(figure 2) suggests it took long time to develop superhumps
in this system.
It also took 6~d ($\sim$60 cycles) since the outburst detection
to fully develop stage B superhumps based on
the $O-C$ diagram.
Since this object has relatively frequent outbursts,
it it likely that the Case A outburst of
\citet{osa03DNoutburst} classification should occur
in this object. However, the observed delay in superhumps 
likely reflects the long growth time of superhumps
(just like the Case B outburst for low-$q$ system;
in most ordinary SU UMa-type dwarf
novae, the growing stage of superhumps is rarely
recorded a few days after the outburst detection,
e.g. \cite{Pdot}).
This delay of appearance of stage B superhumps
is unusually long for an ordinary SU UMa-type dwarf nova and
is even comparable to extreme WZ Sge-type objects
(cf. \cite{kat15wzsge}).

   After reaching the maximum amplitude, the superhump
period became short as in stage B superhumps in
ordinary SU UMa-type dwarf novae
(\cite{Pdot}; \cite{kat13qfromstageA}).  We identified
40$\le E \le$98 as stage B and obtained a mean period
of 0.10541(4)~d and a period derivative
$P_{\rm dot} = \dot{P}/P$ of $+20.8(2.0) \times 10^{-5}$.
The $O-C$ analysis indicates that the times of 
superhumps for $E \ge$106 can be very well expressed
by a period of 0.10444(5)~d, which we consider
the period of stage C superhumps.

\subsection{Mass ratio from stage A superhumps}\label{sec:massratio}

   As described in subsection \ref{sec:superhump},
the modern method using stage A superhumps gives
a very large mass ratio of $q$=0.34(2).
Since the early part of the observations was not
ideally obtained, we give a firmly determined lower limit
of the period of stage A superhumps (0.1075~d),
which corresponds to $q \ge$0.26.
These lower limit is close to the borderline
($q=$0.24) of the development of superhumps
in 3-D numerical simulation \citep{smi07SH}.
Our best value is close to the limit $q\sim$0.33
to develop the 3:1 resonance under condition
of reduced mass-transfer \citep{mur00SHintermediateq}.
Although our observation suffered from uncertainty
due to the gap in the observation, we have demonstrated
that stage A superhump method is applicable to
systems close to the stability limit.

\subsection{Mass ratio and disk radius from stage C superhumps}\label{sec:massratioradiusC}

   As described in \citet{kat13qfromstageA}, the precession
rate of stage C superhumps can be used to estimate
the disk radius if the mass ratio is known, since
the pressure effect can be neglected in cool
post-superoutburst disks.  If the disk radius can be
estimated, we can constrain the mass ratio
(e.g. \cite{kat13j1222}).
The measured $\epsilon^*$ for stage C superhumps of
V1006 Cyg is 0.0518(10).
This value corresponds to a disk radius of 0.37$A$,
where $A$ is the binary separation, for $q$=0.26
and 0.34$A$ for $q$=0.34.

\subsection{Stage B-C transition and rebrightening}\label{sec:stageBC}

   The stage B-C transition described in subsection
\ref{sec:superhump} occurred during the rapid decline
from the superoutburst plateau.  This feature is
different from the behavior from the one in ``textbook''
SU UMa-type dwarf novae, in which stage B-C transition
usually occurs during the later part of the superoutburst
plateau and is usually associated with a small brightening
trend \citep{Pdot}.  In V1006 Cyg, the object instead
faded and a rebrightening was recorded after
the transition.  Six days after this rebrightening,
the object underwent another outburst.  As judged from
the subsequent behavior (E. Pavlenko et al. in preparation),
this outburst was the first normal outburst of the regular
cycle of normal outbursts.\footnote{
   The identification as a rebrightening is also based
   on the similarity of the behavior with the SU UMa-type
   dwarf nova QZ Ser in the period gap (T. Ohshima et al.
   in prep.), which shows only very infrequent outbursts.
}

   The origin of stage B-C transition is still poorly
understood.  In the present case, it appears that
the cooling front started before
the termination of the plateau phase, since the first
rebrightening occurred only three days after the
rapid fading.  Although such early occurrence of
a rebrightening is rarely met in ordinary SU UMa-type
dwarf novae, similar one was observed in the long-$P_{\rm orb}$
system MN Dra (see figure 3 in \cite{ant02var73dra}).
We propose that the mass ratio close to the stability
border of the 3:1 resonance in V1006 Cyg made it
difficult to maintain the tidal instability, and
the thermal and tidal instabilities decoupled
as proposed for ER UMa-type objects and WZ Sge-type
rebrightenings as suggested by \citet{hel01eruma}.
Although \citet{hel01eruma} considered that small $q$
is responsible for this phenomenon, we can extend
the same discussion to objects with large $q$ close
to the stability border of the 3:1 resonance.

   It would be worth mentioning that stage B-C transition
is not usually observed in WZ Sge-type objects
\citep{kat15wzsge}.  It is possible that rapid fading
from the superoutburst plateau (often seen as a temporary
dip) in WZ Sge-type dwarf novae has the same properties
as in V1006 Cyg.  The common behavior (termination of
the plateau phase before appearance of stage C superhumps,
dip-like fading and rebrightening) in WZ Sge-type objects
and objects having mass ratios close to the stability border
may be understood in a unified way:
the small effect of the small tidal torque is
unable to maintain the hot state when the cooling front starts.

\subsection{Comparison with TU Mensae}\label{sec:tumen}

   Up to this work, TU Men was the only established dwarf nova
that shows three types of outbursts (normal, long normal and
superoutbursts; the names here are given in modern sense)
(\cite{war95suuma}; \cite{bat00tumen}).
The only other possible example is
NY Ser \citep{pav14nyser} which showed outbursts with
intermediate durations without superhumps.
Since there were outbursts in V1006 Cyg lasting more than
six days (2007) and more than five days (2009) without superhumps
\citep{pav14nyser}, the present detection of a genuine
superoutburst makes V1006 Cyg the second case showing three
types of outbursts.  This indicates that $P_{\rm orb}$
above the period gap is not an essential condition for displaying
such behavior.

\subsection{Comparison with other systems}\label{sec:comparison}

   The interpretation that superhumps slowly grow in systems
with mass ratios close to the stability limit was first
presented in \citet{Pdot6} for MN Dra.  Although there still
remains uncertainty about $P_{\rm orb}$ of MN Dra,
and the identification of superhumps stages remained somewhat
unclear, the present detection of growing superhumps in
V1006 Cyg has established this interpretation.
In table 1, we list the objects having
long orbital (or superhump) periods and long-lasting stage A
superhumps.  The suspected orbital periods for MN Dra and
CRTS J214738.4$+$244554 are taken from \citet{pav10mndra}
and \citet{Pdot4}, respectively.  By using this period,
we could also obtain the mass ratio for CRTS J214738.4$+$244554
from stage A superhumps.  All the obtained mass ratios
are in the range of 0.20--0.34, consistent with the theoretical
stability limit.

\section*{Acknowledgments}

This work was supported by the Grant-in-Aid
``Initiative for High-Dimensional Data-Driven Science through Deepening
of Sparse Modeling'' (25120007) from the Ministry of Education,
Culture, Sports, Science and Technology (MEXT) of Japan.
This work also was partially supported by grants
of RFBR 15-32-50920 and 15-02-06178,
14-02-00825 and by the VEGA grant No. 2/0002/13.

\section*{Supporting information}

Additional supporting information can be found in the online version
of this article: Figure 2, Tables 1, 2, 3. \\
Supplementary data is available at PASJ Journal online
({\it included at the end in this astro-ph version}).

\newpage

\setcounter{figure}{1}

\setcounter{table}{0}

\begin{figure*}
  \begin{center}
    \FigureFile(140mm,150mm){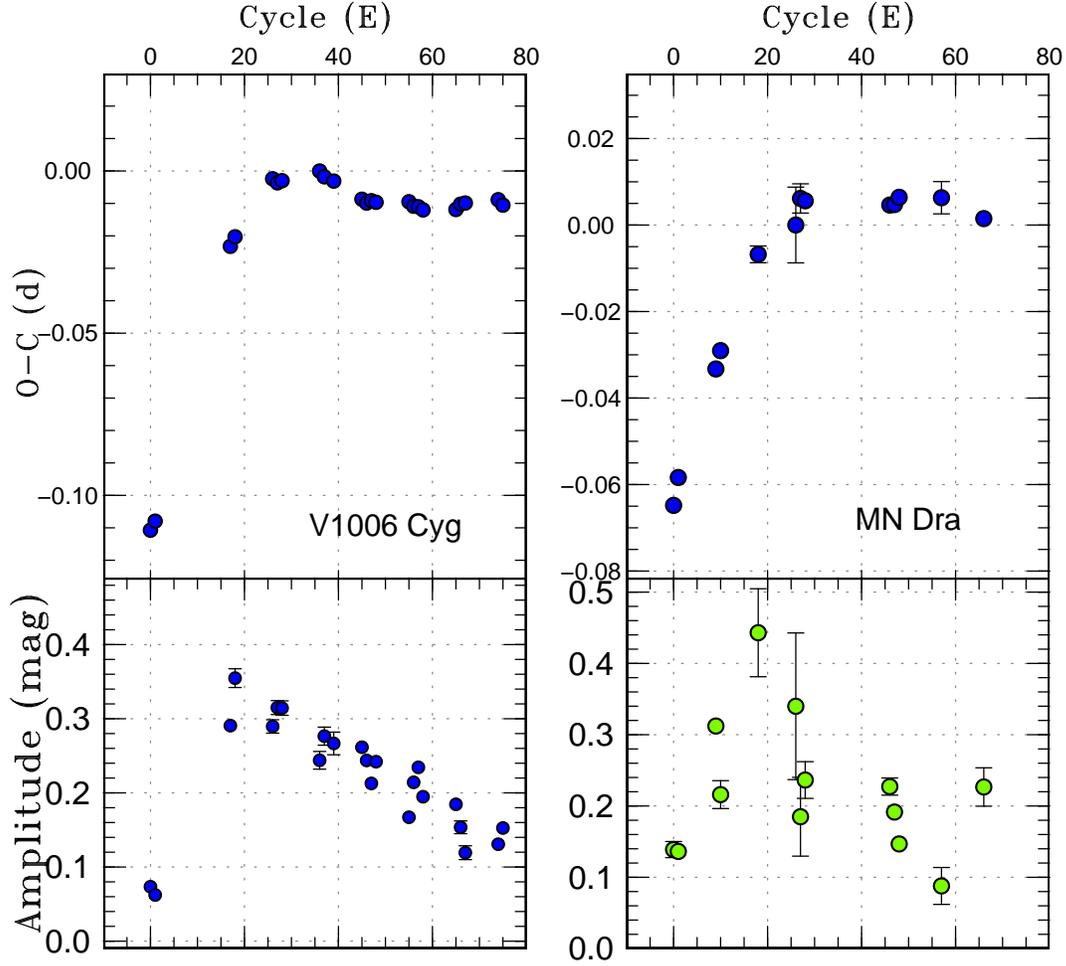}
  \end{center}
  \caption{Comparison of $O-C$ diagrams and variation
  of superhump amplitudes in V1006 Cyg and MN Dra.
  The data for MN Dra are taken from \citet{Pdot6}.
  The behavior of $O-C$ diagrams are almost the same
  in both systems.  In MN Dra with denser observations
  in the early part, the slow growth of the superhump
  amplitudes for two nights were recorded.  Based on
  the similarlity of the $O-C$ diagrams, we consider
  that the superhump amplitudes also grew slowly in
  V1006 Cyg.
  }
  \label{fig:v1006cygmndracomp}
\end{figure*}

\begin{table*}
\caption{Comparison of SU UMa-type objects with long phase of stage A superhumps}\label{tab:longPsuuma}
\begin{center}
\begin{tabular}{cccccccc}
\hline
Object & $P_{\rm orb}$\commenta & $P_{\rm A}$\commentb
       & $P_{\rm B}$\commentc & $P_{\rm C}$\commentd
       & dur\commente & $q$\commentf & References \\
\hline
V1006 Cyg (2015) & 0.09903(9) & 0.1093(3) & 0.10541(4) & 0.10444(5) & $\ge$32 & 0.34(2) & This work \\
MN Dra (2012) & 0.0998(2) & 0.10993(9) & 0.10530(6) & -- & $\ge$39 & 0.327(5) & {Kato} et~al. (2014) \\
MN Dra (2013) & 0.0998(2) & 0.1082(1) & 0.10504(7) & -- & $\ge$18 & 0.258(5) & {Kato} et~al. (2014) \\
CRTS J214738.4$+$244554 (2011) & 0.09273(3) & 0.0992(3) & 0.09715(2) & -- & $\ge$21 & 0.204(11) & {Kato} et~al. (2015) \\
OT J064833.4$+$065624 (2014) & -- & 0.1052(4) & 0.10033(3) & -- & $\ge$38 & -- & {Kato} et~al. (2015) \\
\hline
  \multicolumn{8}{l}{\commenta Orbital period (d).} \\
  \multicolumn{8}{l}{\commentb Period of stage A superhumps (d).} \\
  \multicolumn{8}{l}{\commentc Period of stage B superhumps (d).} \\
  \multicolumn{8}{l}{\commentd Period of stage C superhumps (d).} \\
  \multicolumn{8}{l}{\commente Duration of stage A (cycles).} \\
  \multicolumn{8}{l}{\commentf Determined from stage A superhumps.} \\
\end{tabular}
\end{center}
\end{table*}

\begin{table*}
\caption{Superhump maxima of V1006 Cyg (2015)}
\begin{center}
\begin{tabular}{rp{55pt}p{40pt}r@{.}lr}
\hline
\multicolumn{1}{c}{$E$} & \multicolumn{1}{c}{max\commenta} & \multicolumn{1}{c}{error} & \multicolumn{2}{c}{$O-C$\commentb} & \multicolumn{1}{c}{$N$\commentc} \\
\hline
0 & 57219.4523 & 0.0008 & $-$0&1077 & 156 \\
1 & 57219.5605 & 0.0011 & $-$0&1046 & 106 \\
17 & 57221.3318 & 0.0001 & $-$0&0155 & 288 \\
18 & 57221.4401 & 0.0006 & $-$0&0123 & 42 \\
26 & 57222.3013 & 0.0004 & 0&0077 & 120 \\
27 & 57222.4055 & 0.0004 & 0&0068 & 72 \\
28 & 57222.5115 & 0.0004 & 0&0077 & 71 \\
36 & 57223.3578 & 0.0007 & 0&0129 & 69 \\
37 & 57223.4614 & 0.0006 & 0&0113 & 55 \\
39 & 57223.6709 & 0.0006 & 0&0106 & 49 \\
45 & 57224.2978 & 0.0004 & 0&0066 & 58 \\
46 & 57224.4020 & 0.0003 & 0&0057 & 339 \\
47 & 57224.5081 & 0.0003 & 0&0067 & 393 \\
48 & 57224.6131 & 0.0004 & 0&0065 & 104 \\
55 & 57225.3511 & 0.0005 & 0&0086 & 100 \\
56 & 57225.4551 & 0.0004 & 0&0075 & 223 \\
57 & 57225.5604 & 0.0004 & 0&0076 & 175 \\
58 & 57225.6648 & 0.0006 & 0&0069 & 72 \\
65 & 57226.4028 & 0.0004 & 0&0089 & 167 \\
66 & 57226.5099 & 0.0008 & 0&0108 & 153 \\
67 & 57226.6156 & 0.0013 & 0&0114 & 72 \\
74 & 57227.3545 & 0.0008 & 0&0144 & 74 \\
75 & 57227.4583 & 0.0007 & 0&0130 & 73 \\
83 & 57228.3059 & 0.0009 & 0&0195 & 45 \\
84 & 57228.4106 & 0.0007 & 0&0191 & 97 \\
85 & 57228.5169 & 0.0008 & 0&0202 & 113 \\
93 & 57229.3629 & 0.0005 & 0&0251 & 55 \\
94 & 57229.4677 & 0.0010 & 0&0248 & 44 \\
102 & 57230.3062 & 0.0011 & 0&0222 & 52 \\
103 & 57230.4096 & 0.0012 & 0&0204 & 126 \\
104 & 57230.5149 & 0.0010 & 0&0206 & 86 \\
122 & 57232.3943 & 0.0010 & 0&0075 & 56 \\
123 & 57232.5043 & 0.0014 & 0&0124 & 57 \\
125 & 57232.7077 & 0.0006 & 0&0055 & 539 \\
126 & 57232.8108 & 0.0006 & 0&0035 & 531 \\
131 & 57233.3405 & 0.0018 & 0&0074 & 43 \\
132 & 57233.4475 & 0.0018 & 0&0093 & 140 \\
133 & 57233.5574 & 0.0017 & 0&0141 & 36 \\
137 & 57233.9733 & 0.0019 & 0&0094 & 24 \\
140 & 57234.2937 & 0.0009 & 0&0145 & 53 \\
141 & 57234.3891 & 0.0010 & 0&0047 & 57 \\
142 & 57234.4874 & 0.0023 & $-$0&0021 & 57 \\
160 & 57236.3651 & 0.0009 & $-$0&0169 & 121 \\
161 & 57236.4683 & 0.0011 & $-$0&0188 & 109 \\
170 & 57237.4130 & 0.0025 & $-$0&0204 & 110 \\
198 & 57240.3309 & 0.0015 & $-$0&0464 & 87 \\
199 & 57240.4300 & 0.0024 & $-$0&0523 & 115 \\
200 & 57240.5527 & 0.0014 & $-$0&0349 & 69 \\
\hline
  \multicolumn{6}{l}{\commenta BJD$-$2400000.} \\
  \multicolumn{6}{l}{\commentb Against max $= 2457219.5600 + 0.105138$.} \\
  \multicolumn{6}{l}{\commentc Number of points used to determine the maximum.} \\
\end{tabular}
\end{center}
\end{table*}

\begin{table*}
\caption{Log of observations of V1006 Cyg (2015)}
\begin{center}
\begin{tabular}{cccccccccc}
\hline
Start\commenta & End\commenta & $N$\commentb & Code\commentc & filter\commentd &
Start\commenta & End\commenta & $N$\commentb & Code\commentc & filter\commentd \\
\hline
57219.0237 & 57219.2394 &  405 &  Mdy &  V & 57231.7318 & 57231.8181 &  484 &  LCO &  C \\
57219.3980 & 57219.4864 &   67 &  Trt &  C & 57232.2778 & 57232.5673 &  196 &  CRI &  V \\
57219.4226 & 57219.6117 &  249 &  RPc &  V & 57232.6069 & 57232.8478 & 1163 &  LCO &  C \\
57221.2790 & 57221.3793 &   87 &  CRI &  I & 57232.7119 & 57232.9519 &  350 &  COO &  C \\
57221.2793 & 57221.3796 &   87 &  CRI &  R & 57233.2564 & 57233.5640 &  195 &  CRI &  V \\
57221.2796 & 57221.3787 &   86 &  CRI &  B & 57233.3993 & 57233.4743 &   94 &  RPc &  V \\
57221.2799 & 57221.3790 &   85 &  CRI &  V & 57233.8786 & 57234.0075 &   39 &  COO &  C \\
57221.4283 & 57221.4774 &   42 &  IMi &  V & 57234.2623 & 57234.5547 &  198 &  CRI &  V \\
57222.2065 & 57222.2836 &  113 &  Ioh &  C & 57235.2832 & 57235.5584 &  183 &  CRI &  V \\
57222.2594 & 57222.5643 &  280 &  CRI &  V & 57235.3232 & 57235.5705 &  165 &  DPV &  C \\
57223.2543 & 57223.5649 &  292 &  CRI &  V & 57235.3672 & 57235.5476 &  124 &  Shu &  C \\
57223.6402 & 57223.6791 &   49 &  deM &  C & 57235.4643 & 57235.6187 &   13 &  COO &  C \\
57224.2714 & 57224.5465 &  230 &  CRI &  V & 57236.2541 & 57236.5254 &  125 &  CRI &  V \\
57224.3371 & 57224.5392 &  260 &  DPV &  C & 57236.3156 & 57236.5207 &  177 &  Shu &  C \\
57224.3705 & 57224.6742 &  359 &  deM &  C & 57237.3182 & 57237.5208 &  143 &  Shu &  C \\
57224.4079 & 57224.5132 &  128 &  RPc &  V & 57237.3766 & 57237.5597 &   72 &  CRI &  V \\
57224.4289 & 57224.5236 &  124 &  IMi &  V & 57238.2883 & 57238.5461 &  103 &  CRI &  V \\
57225.2640 & 57225.5473 &  191 &  CRI &  V & 57238.3035 & 57238.3228 &   17 &  Shu &  C \\
57225.3599 & 57225.4483 &  114 &  DPV &  C & 57238.3495 & 57238.5317 &  123 &  DPV &  C \\
57225.4328 & 57225.6789 &  301 &  deM &  C & 57238.3909 & 57238.5692 &  180 &  RPc &  C \\
57225.4685 & 57225.5695 &  113 &  Trt &  V & 57238.4168 & 57238.5540 &  192 &  Trt &  V \\
57226.3121 & 57226.5320 &  276 &  DPV &  C & 57239.1894 & 57239.2957 &  279 &  KU1 &  C \\
57226.3679 & 57226.4597 &   73 &  CRI &  V & 57239.2470 & 57239.5708 &  178 &  CRI &  V \\
57226.4991 & 57226.6241 &  174 &  RPc &  V & 57239.2983 & 57239.5495 &   79 &  Shu &  V \\
57227.2796 & 57227.5517 &  240 &  CRI &  V & 57239.3176 & 57239.4514 &   98 &  Trt &  V \\
57228.2790 & 57228.5579 &  186 &  CRI &  V & 57240.0578 & 57240.1351 &  193 &  KU1 &  C \\
57228.3883 & 57228.5653 &  118 &  DPV &  C & 57240.2694 & 57240.5664 &  201 &  CRI &  V \\
57229.2688 & 57229.5634 &  181 &  CRI &  V & 57240.3250 & 57240.5372 &  144 &  DPV &  C \\
57230.2753 & 57230.5615 &  127 &  CRI &  V & 57241.2502 & 57241.5681 &  214 &  CRI &  V \\
57230.3108 & 57230.5230 &  144 &  DPV &  C & 57241.3033 & 57241.3973 &   65 &  DPV &  C \\
57230.3801 & 57230.4999 &   58 &  Trt &  V & 57241.3375 & 57241.4990 &   43 &  Kaz &  C \\
57231.3481 & 57231.5638 &  100 &  CRI &  V & 57242.2957 & 57242.3778 &   56 &  DPV &  C \\

\hline
  \multicolumn{10}{l}{\commenta JD$-$2400000.} \\
  \multicolumn{10}{l}{\commentb Number of observations.} \\
  \multicolumn{10}{l}{\parbox{400pt}{\commentc Key to observers:
COO (L. Cook),
CRI (Crimean Astrophys. Obs.),
DPV (P. Dubovsky),
deM (E. de Miguel),
IMi (I. Miller),
Ioh (H. Itoh),
KU (Kyoto U., campus obs.),
Kaz (Kazan' Univ. Obs.),
LCO (C. Littlefield),
Mdy (Y. Maeda),
RPc (R. Pickard),
Shu (S. Shugarov team),
Trt (T. Tordai).
  }} \\
  \multicolumn{10}{l}{\commentd The filter name C represents unfiltered observations.} \\
\end{tabular}
\end{center}
\end{table*}

\end{document}